\documentclass[twocolumn,showpacs,amsfonts,aps,prc,floatfix,%
superscriptaddress,nofootinbib]{revtex4-1} 

\usepackage{amsmath}
\usepackage{bm}
\usepackage{graphicx}
\usepackage{url}
\usepackage{longtable}

\newcommand{\be}[1]{\begin{equation}\label{#1}}
\newcommand{\ee}{\end{equation}}

\newcommand{\vtwo}{$v_2$ }

\newcommand{\lsim}{\mbox{\raisebox{-0.6ex}{$\stackrel{<}{\sim}$}}\:}

\usepackage{color}

\begin{document}


\title{Elliptic flow in U+U collisions at $\sqrt{s_{NN}}$ = 
200 GeV 
and in Pb+Pb collisions\\
at $\sqrt{s_{NN}}$ = 2.76 TeV:
Prediction from a hybrid approach
}
\date{\today}

\author{Tetsufumi Hirano}
\email{hirano@phys.s.u-tokyo.ac.jp}
\affiliation{Department of Physics, The University of Tokyo,
Tokyo 113-0033, Japan}
\affiliation{Nuclear Science Division, Lawrence Berkeley National Laboratory,
Berkeley, CA 94720, USA}

\author{Pasi Huovinen}
\email{huovinen@th.physik.uni-frankfurt.de}
\affiliation{Institut f\"ur Theoretische Physik, Johann Wolfgang Goethe-Universit\"at, 60438 Frankfurt am Main, Germany  
}

\author{Yasushi Nara}
\email{nara@aiu.ac.jp}
\affiliation{Akita International University, Yuwa, Akita-city 010-1292, Japan}

\begin{abstract}
We predict the elliptic flow parameter \vtwo
in U+U collisions at $\sqrt{s_{NN}}=200$ GeV
and in Pb+Pb collisions at
$\sqrt{s_{NN}}$ = 2.76 TeV
using a hybrid model in which the evolution of the quark gluon
  plasma is described by ideal hydrodynamics with a state-of-the-art
  lattice QCD equation of state, and the subsequent hadronic stage by
  a hadron cascade model.

\end{abstract}
\pacs{25.75.-q, 25.75.Nq, 12.38.Mh, 12.38.Qk}

\maketitle


One of the major discoveries at 
Relativistic Heavy Ion Collider (RHIC)
in Brookhaven National Laboratory (BNL)
was that 
for the first time in relativistic heavy ion collisions, the
elliptic flow
appeared 
to be as large as an ideal hydrodynamic prediction \cite{BNL}.
Since viscosity and any other dissipative effects
vanish in ideal hydrodynamics,
and tiny viscosity requires
a strong coupling of constituents
(quarks and gluons in our case),
this discovery established the 
new paradigm
of strongly coupled quark gluon plasma~\cite{Gyulassy:2004vg,sQGP}.

In noncentral collisions,
rescatterings of the created particles convert the initial
  spatial anisotropy of the reaction zone to anisotropic particle distribution
 \cite{Ollitrault}.
Ideal hydrodynamics predicts that the ratio of these anisotropies is 
$v_2/\varepsilon \sim $0.2 
almost independent of centrality at the
RHIC energies \cite{Kolb:2000sd}.
Here $v_{2}$ is the second Fourier coefficient
of the azimuthal
distribution of final particles,
and $\varepsilon$ is the initial eccentricity
of the produced matter.
On the other hand, 
in the dilute regime kinetic theory predicts \vtwo to be
  proportional to the particle multiplicity per unit rapidity,
  $dN/dy$ \cite{Heiselberg:1998es,Kolb:2000fha}.
Thus, the response of the system, $v_2/\varepsilon$,
provides information about the transport properties
of the QGP.
Experimentally $v_2/\varepsilon$ is seen to increase with increasing
transverse density $(1/S)dN/dy$ \cite{Adler:2002pu,Alt:2003ab},
where $S$ is the transverse area of the collision zone, 
until it reaches the so-called
hydrodynamic limit, $v_{2}/\varepsilon \sim 0.2$,
in central Au+Au collisions at RHIC.
With the agreement of the hydrodynamical prediction of the
  particle mass dependence of $v_2(p_T)$~\cite{Huovinen:2001cy} with
  the data~\cite{Adams:2003am,Adler:2003kt}, this is considered as
  evidence for the discovery of the perfect-fluid nature of the QGP
  \cite{BNL}.

After observing the increase of $v_2/\varepsilon$ with increasing
  transverse density, it is natural to ask what happens if the
  transverse density increases beyond that achieved at
  RHIC~\cite{Heinz:2004ir}. Will it saturate to the value observed at
  RHIC, as expected if the system behaves like a perfect fluid,
  or will it keep increasing?
One suggested way to extend the transverse density is
to perform uranium-uranium collisions \cite{Heinz:2004ir}.
Since uranium nuclei are deformed
and larger than gold nuclei,
one can expect large transverse density with
finite eccentricity
in the body-body collisions
at vanishing impact parameter\footnote{
The idea of collisions of deformed nuclei
is not new and one can find
literature on this subject. See, e.g., Refs.~\cite{Kolb:2000sd, Nonaka:2000ek}.
}.
Some Monte Carlo studies
show that, even though one cannot control
the orientation of colliding nuclei,
events with high multiplicity though finite
eccentricity can be selected
in the usual triggering process
\cite{Heinz:2004ir,Kuhlman:2005ts,Nepali:2006ep,Nepali:2007an,Masui:2009qk}.
Another way to extend the transverse energy
is to increase the collision energy to generate
more particles in collisions.
This is going to happen very soon in the Large Hadron Collider (LHC)
heavy ion program.
In this Rapid Communication we predict elliptic flow parameters
  both in U+U collisions at RHIC and Pb+Pb collisions at LHC using
  a hybrid model based on ideal hydrodynamics and hadron cascade.


We describe space-time evolution of the QGP
by ideal hydrodynamics \cite{Hirano:2001eu}
with the recent
lattice QCD equation of state \cite{Huovinen:2009yb}.
After expansion and cooling,
the system turns into hadronic matter.
We switch from hydrodynamics to a kinetic approach 
at a switching temperature $T_{\mathrm{sw}}$
and employ a hadronic cascade model, JAM \cite{JAM}, to describe
the subsequent space-time evolution
of hadronic matter.

Our EoS, $s95p$-v1.1, is a slightly modified version of the
$s95p$-v1 EoS presented in~\cite{Huovinen:2009yb}. It interpolates
between hadron resonance gas at low temperatures, and recent lattice
QCD results by the hotQCD
collaboration~\cite{Cheng:2007jq,Bazavov:2009zn} at high temperatures
in the same way as $s95p$-v1, but the hadron resonance gas part 
contains the same hadrons and resonances as the JAM hadron
cascade~\cite{JAM}. The details of the interpolating procedure are
explained in~\cite{Huovinen:2009yb} and the parametrization and EoS
tables are available at~\cite{EoSsite}.

\begin{figure*}[htb]
\includegraphics[width=3.4in]{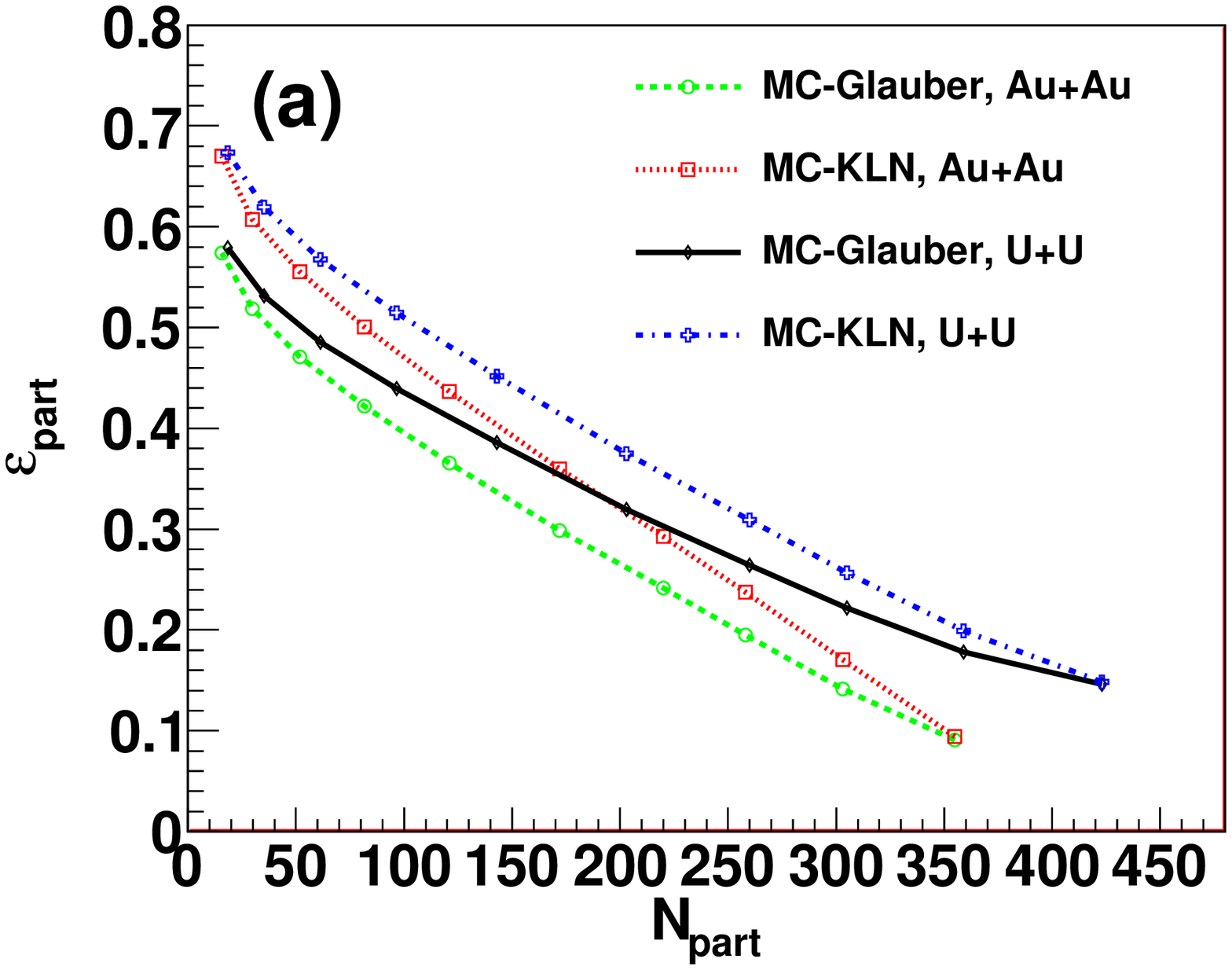}
\includegraphics[width=3.4in]{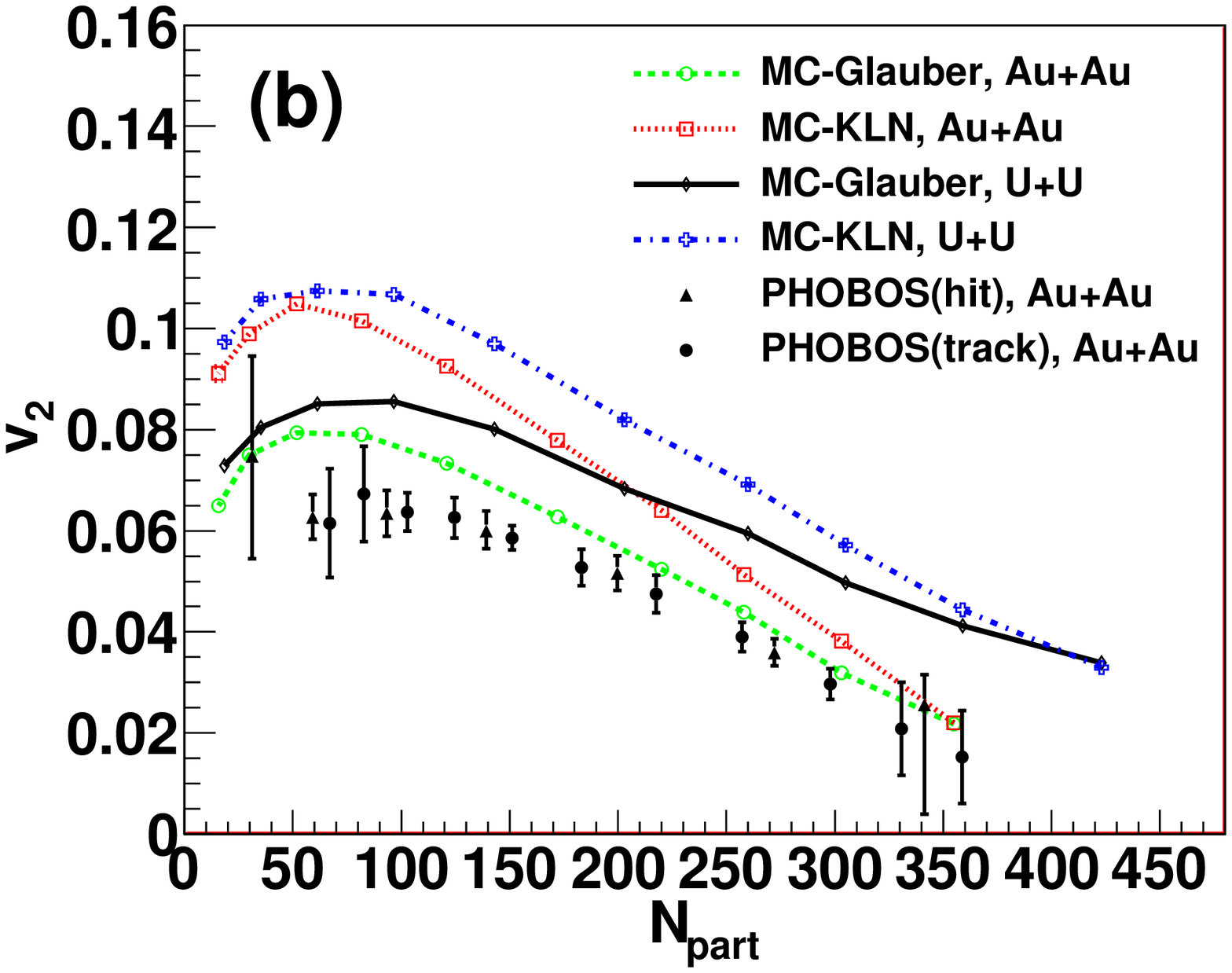}
\caption{(Color online)
(a) Initial state eccentricity $\varepsilon_\mathrm{part}$ and 
(b) $v_2$ as a function of $N_{\mathrm{part}}$ 
in Au+Au and U+U
collisions at $\sqrt{s_{NN}}$ = 200 GeV.
Experimental data of $v_2$ in Au+Au collisions
are obtained by PHOBOS Collaboration \cite{Back:2004mh}.
}
\label{fig:AuU}
\end{figure*}
%

For initial conditions, we employ two Monte-Carlo
approaches to simulate
collisions of two energetic nuclei: Monte-Carlo Glauber (MC-Glauber) model
\cite{Miller:2007ri} 
and Monte-Carlo Kharzeev-Levin-Nardi (MC-KLN) model~\cite{MCKLN}.
In the MC-Glauber model,
one calculates the number of participants $N_{\mathrm{part}}$ and 
the number of binary collisions $N_{\mathrm{coll}}$
for a given nuclear density distribution.
We model the initial entropy distribution
in hydrodynamic simulations
as a linear combination
of the number distribution of participants $\rho_{\mathrm{part}}=\frac{dN_{\mathrm{part}}}{d^{2}\bm{x}_{\perp}}$
and that of binary collisions $\rho_{\mathrm{coll}}=\frac{dN_{\mathrm{coll}}}{d^{2}\bm{x}_{\perp}}$
in the transverse plane:
\begin{eqnarray}
\frac{dS}{d^{2}\bm{x}_{\perp}} & \propto &
\frac{1-\alpha}{2}\rho_{\mathrm{part}}(\bm{x}_{\perp})
+ \alpha \rho_{\mathrm{coll}}(\bm{x}_{\perp}).
\label{eq:dsdx2}
\end{eqnarray}
We generate the number distributions in an event-by-event basis,
align them
to match the main and sub axes of the ellipsoids,
and average over many events for a given centrality bin
to obtain a smooth distribution \cite{Hirano:2009ah}.
The eccentricity of the initial profile is then evaluated with respect 
to participant plane, $\varepsilon_{\mathrm{part}}$~\cite{Alver:2008zza}.
We do the centrality cuts according to the $N_{\mathrm{part}}$ distribution
from the MC-Glauber model instead of using the optical Glauber limit as
was done in Ref.~\cite{Hirano:2009ah}.
The free parameters of the model, the mixing parameter $\alpha = 0.18$
and the proportionality constant in Eq.~(\ref{eq:dsdx2}), are chosen
to reproduce transverse momentum
spectra for pions, kaons, and protons
from central (0-5\%) to peripheral (70-80\%)
events in Au+Au collisions at $\sqrt{s_{NN}} = 200$ GeV
obtained by the PHENIX Collaboration \cite{Adler:2003cb}. 
We also choose the switching temperature as $T_{\mathrm{sw}} = 155$ MeV
to describe the relative yields for pions, kaons, and protons in these
data.

MC-KLN model is a Monte Carlo version of
the factorized Kharzeev-Levin-Nardi (fKLN) model~\cite{fKLN}.
In the MC-KLN model, gluon production is
obtained by numerical integration of the $k_t$-factorized
formula~\cite{KLN} at each transverse grid.
The fluctuation of gluon distribution
due to the position of hard sources
(nucleons) in the transverse plane is
taken into account in MC-KLN.
Using the thickness function $T_{A}$,
we parametrize the saturation scale for a nucleus A as
\begin{equation}
Q_{s,A}^2 (x; \bm{x}_\perp)  =  2\ \text{GeV}^2
\frac{T_A(\bm{x}_\perp)}{1.53\ \text{fm}^{-2}}
\left(\frac{0.01}{x}\right)^{\lambda}
\label{eq:qs2}
\end{equation}
and similarly for a nucleus $B$.
We choose $\lambda=0.28$ and a proportionality constant
in the unintegrated gluon distribution in the $k_{t}$-factorized formula
to reproduce centrality dependence of $p_{T}$ spectra
for pions, kaons, and protons as above.

Using the same parameter set as above, we
calculate initial entropy distribution in U+U collisions
by changing the nuclear density from
 gold to uranium.
To take account of the prolate deformation of
uranium nuclei,
we parametrize the
radius parameter in the Woods-Saxon distribution
as 
\begin{eqnarray}
R(\theta, \phi) & = & R_{0}\left( 1+\beta_{2}Y_{20}(\theta, \phi)
 + \beta_{4}Y_{40}(\theta, \phi) \right),
\end{eqnarray}
where $Y_{lm}$ is the spherical harmonic function,
$R_{0} = 6.86$ fm,
$\beta_{2} = 0.28$ and $\beta_{4} = 0.093$ \cite{Filip:2009zz}.
Note that to account of
the finite size of nucleons in the Monte Carlo approach,
we have adjusted $R_{0}$ above 
and the diffuseness parameter $\delta r = 0.44$
to retain the nuclear density as in the original Woods-Saxon distribution
\cite{Hirano:2009ah}.
We also take into account
that colliding uranium nuclei are randomly
oriented
in each event.


Figure \ref{fig:AuU} (a) shows
initial eccentricity with respect to participant plane
in Au+Au and U+U collisions at $\sqrt{s_{NN}} = 200$ GeV
as a function of the number of participants.
At each of the ten centrality bins the average eccentricity and
  the average number of participants $\langle N_{\mathrm{part}} \rangle$ 
  were calculated using both the MC-Glauber and the MC-KLN models.
Since the eccentricity is measured in
the participant plane, it is finite
even in the very central (0-5\%) Au+Au collisions.
As previously known,
the MC-KLN model leads to $\sim$20-30\% larger eccentricity than 
the MC-Glauber model except in the most central events 
 \cite{fKLN,Hirano:2005xf}.
In most central 5\% of U+U collisions eccentricity reaches
0.146 in the MC-Glauber model and 0.148 in the MC-KLN model.
The eccentricity is larger in U+U than in Au+Au collisions. 
Due to the deformed shape of uranium nucleus, this holds not only 
at fixed number of participants, but also at fixed centrality.
The difference, however, 
decreases with decreasing centrality, and there is almost 
no difference in the very peripheral events (70-80\%).

In Fig.~\ref{fig:AuU} (b),
$v_2$ 
 in Au+Au collisions is compared with
the \vtwo in U+U collisions.
Since the rule of thumb is that larger eccentricity leads to
larger momentum anisotropy and \vtwo, the systematics of
$v_2(N_{\mathrm{part}})$ is similar to that of
$\varepsilon_{\mathrm{part}}(N_{\mathrm{part}})$: \vtwo is larger in
U+U collisions than in Au+Au collisions, and MC-KLN initialization
leads to larger \vtwo than MC-Glauber initialization. As well,
$v_2$ first increases
with decreasing $N_{\mathrm{part}}$,
which reflects increasing initial eccentricity,
but once $N_{\mathrm{part}}$ falls below $\sim 50$, \vtwo begins 
to decrease. This is due to the short lifetime of the system which 
does not allow the flow to fully build up, and to the large fraction 
of the lifetime spent in the hadronic phase where dissipative effects
are large.

Results from the MC-Glauber initialization
almost reproduce
the PHOBOS data \cite{Back:2004mh} in Au+Au collisions.
This indicates that
there is little room for QGP viscosity
in the model calculations.
On the other hand, apparent discrepancy
between the results from the MC-KLN initialization
and the PHOBOS data
means that 
viscous corrections during the plasma phase are required.

Within the color glass condensate picture, the collision energy
  dependence is taken into account through the saturation scale,
  $Q_s$. This allows us to simulate the Pb+Pb collisions at
  $\sqrt{s_{NN}} = 2.76$ TeV by using the MC-KLN model, adjusting the
  collision energy parameter and the nuclear density parametrization,
  and keeping all the other parameters unchanged. This is a consistent
  way to study the differences between collisions at $\sqrt{s_{NN}} =
  62.4$ GeV and 2.76 TeV energies, but it may be too naive, since the
  MC-KLN model does not take into account running coupling corrections
  to the evolution equation~\cite{Albacete:2007sm}. At RHIC energies these
  effects are known to have only a small effect, but at LHC they lead
  to a clearly lower multiplicity~\cite{Albacete:2007sm,mckt}. On the
  other hand, these effects hardly affect the
  eccentricity~\cite{mckt}, which allows us to study the effects of
  the uncertainty in the final particle multiplicity simply by
  adjusting the overall factor in
  the unintegrated gluon distribution function.
  Our default approach is to use the MC-KLN model
  with the same factor 
  than in the RHIC calculations. In 5\% most
  central Pb+Pb collisions this leads to multiplicity
  $dN_{\mathrm{ch}}/d\eta \sim 1600$ at midrapidity ($\mid \eta \mid <
  1$). We also reduce the factor to obtain multiplicities
  $dN_{\mathrm{ch}}/d\eta \sim $1400 (set 1), as predicted in
  Ref.~\cite{Albacete:2007sm}, and $\sim $1200 (set 2)\footnote{ $\eta$ is
  not the shear viscous coefficient but the
  pseudorapidity.  }.

%
 \begin{figure*}[htb]
\includegraphics[width=3.4in]{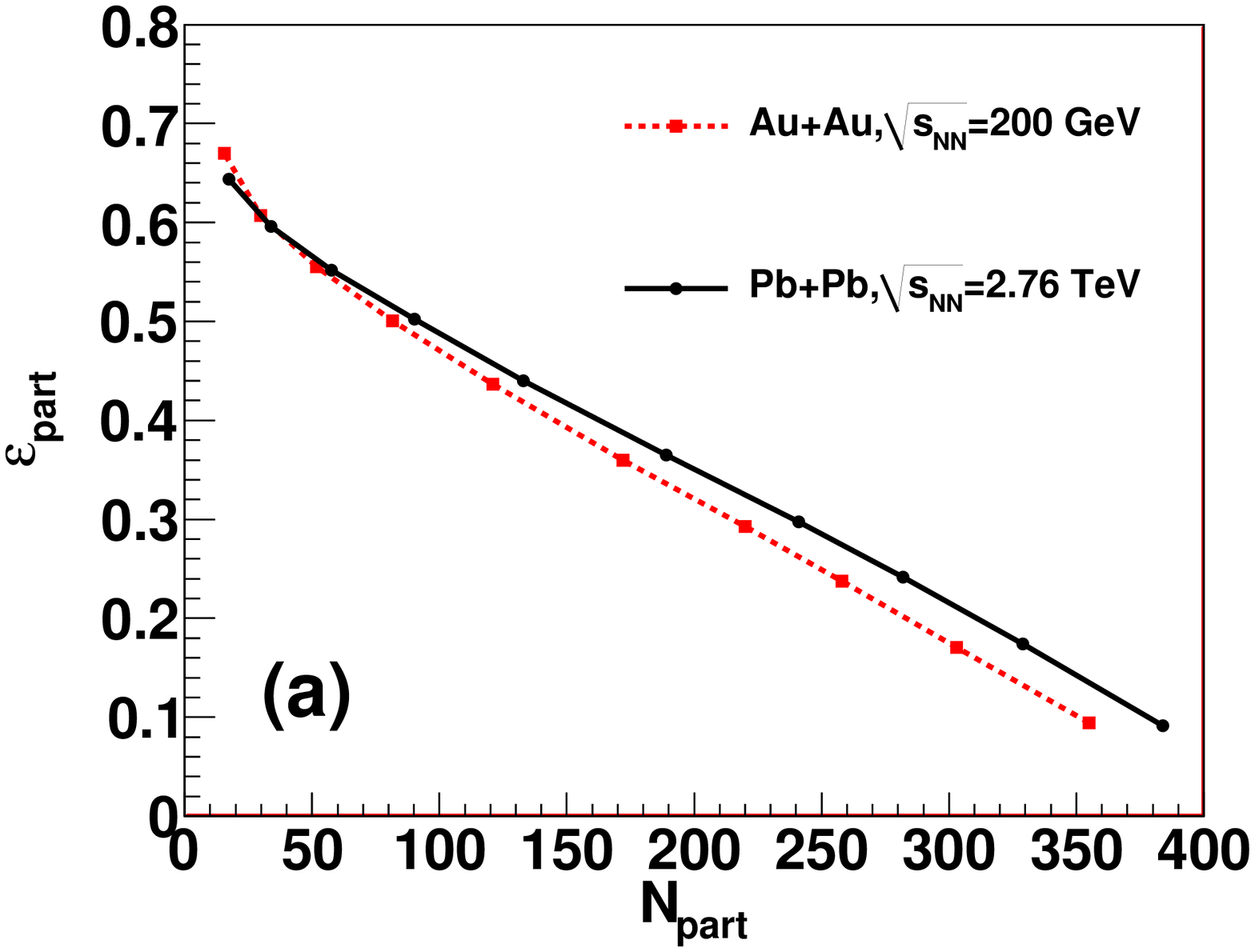}
\includegraphics[width=3.4in]{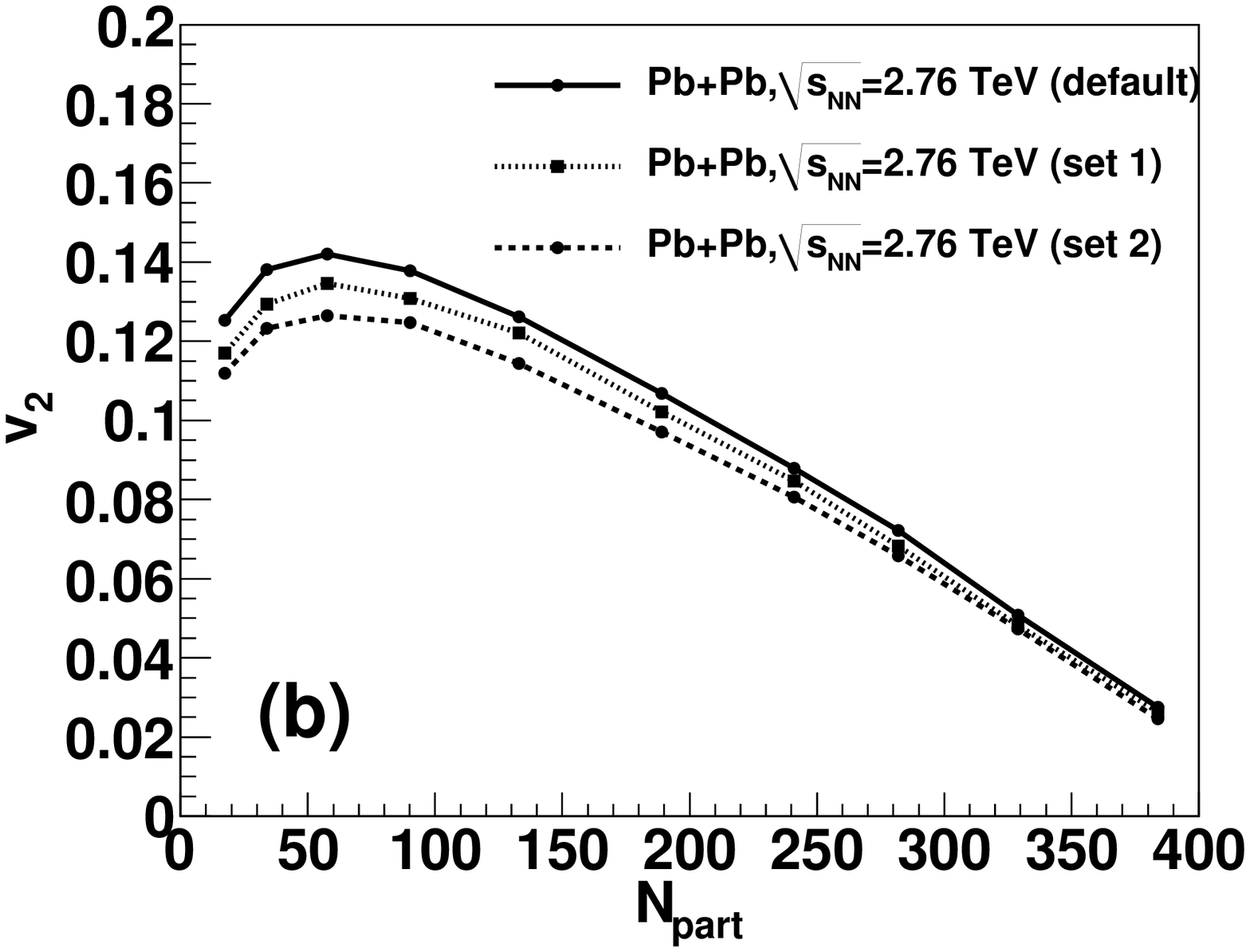}
\caption{(Color online)
(a) $\varepsilon_{\mathrm{part}}$ as a function of $N_{\mathrm{part}}$ 
in Au+Au collisions at $\sqrt{s_{NN}}$ = 200 GeV (dashed)
and in Pb+Pb collisions at $\sqrt{s_{NN}}$ = 2.76 TeV (solid).
(b) $v_2$ as function of number of participants $N_{\mathrm{part}}$ in
Pb+Pb collisions at $\sqrt{s_{NN}}$ = 2.76 TeV for three different
multiplicities in 0-5\% centrality: $dN_{\mathrm{ch}}/d\eta \sim$ 1600 (solid),
1400 (dotted) and 1200 (dashed).
Each point
from right to left corresponds to 0-5, 5-10, 10-15, 15-20,
20-30, 30-40, 40-50, 50-60, 60-70, and 70-80\% centrality, respectively. 
}
 \label{fig:v2nch}
 \end{figure*}
%

Our result for the initial state eccentricity as function of the
  number of participants in Au+Au collisions at $\sqrt{s_{NN}}$ =200
  GeV, and in Pb+Pb collisions at $\sqrt{s_{NN}}$ = 2.76 TeV is shown
  in Fig.~\ref{fig:v2nch} (a). As mentioned, the uncertainty in the
  multiplicity in collisions at $\sqrt{s_{NN}}$ = 2.76 TeV does not
  affect the eccentricity and we show the result obtained using our
  default setting. For a fixed $N_{\mathrm{part}}$, eccentricity at
  LHC is apparently larger than that at RHIC. However, this is due
  solely to the larger size of colliding nuclei.  If one compares the
  eccentricity at a fixed \textit{centrality} (see each point in the
  figure), eccentricities are essentially the same.

In Fig.~\ref{fig:v2nch} (b), $v_2$ in Pb+Pb collisions at
  $\sqrt{s_{NN}}$ = 2.76 TeV is shown as a function of the number of
  participants for three different multiplicities in central
  collisions. The larger the multiplicity, the larger the $v_2$, but
  even at the lowest setting of multiplicity, $v_2$ is clearly larger
  than in the Au+Au collisions at $\sqrt{s_{NN}} = 200$ GeV.

 \begin{figure}[htb]
\includegraphics[width=3.4in]{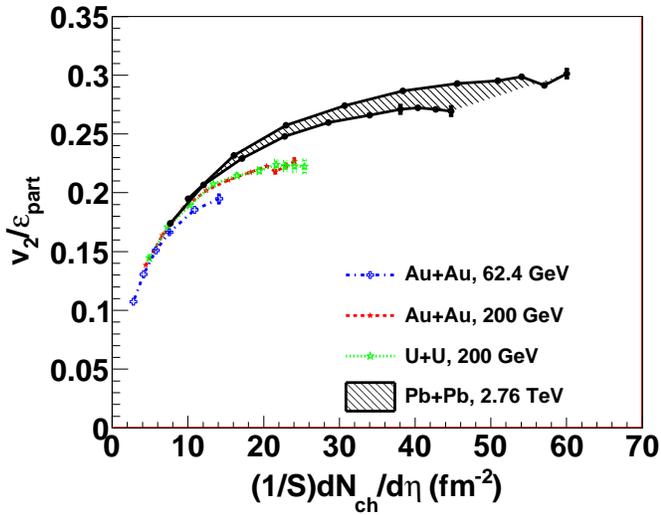}
\caption{$v_2/\varepsilon_{\mathrm{part}}$ as a function of transverse
  density in Au+Au collisions at $\sqrt{s_{NN}}$ = 62.4 (dash-dotted)
  and 200 GeV (dashed), in U+U collisions at $\sqrt{s_{NN}}$ = 200 GeV
  (dotted), and in Pb+Pb collisions at $\sqrt{s_{NN}}$= 2.76 TeV
  (band). The band depicting the Pb+Pb collisions spans the results
  obtained using the multiplicities $1200 < dN_{\mathrm{ch}}/d\eta < 1600$ in
  5\% most central central collisions.}
 \label{fig:v2eccLHC}
 \end{figure}
%

This behavior is clearly visible in Fig.~\ref{fig:v2eccLHC} where we plot
$v_2/\varepsilon_{\mathrm{part}}$ as a function of the transverse
charged particle density $(1/S)dN_{\mathrm{ch}}/d\eta$ at midrapidity
($\mid \eta \mid < 1$) for various collision systems and
energies.  First, as expected, the system in U+U
collisions at $\sqrt{s_{NN}}$ =200 GeV is denser than in Au+Au
collisions at the same energy. At initial time $\tau_{0}$ = 0.6
fm/$c$, the maximum temperature (energy density) in the most central
5\% of U+U collisions is $T_{0}=367$ MeV ($e_{0}=33.4$ GeV/fm$^{3}$)
and $T_{0}=361$ MeV ($e_{0}=31.4$ GeV/fm$^{3}$) in the Au+Au
collisions of the same centrality. This corresponds to charged
particle transverse densities of 25.4 and 24.1, respectively, which
means that the transverse density in U+U collisions is indeed larger,
but only by $\sim$ 6\%\footnote{With sufficient statistics, one may
  make more severe centrality cut (e.g., 0-3\%) to obtain larger
  transverse particle density.  Multiplicity fluctuation in the
  centrality cut, which we do not take into account, could also
  enhance the transverse particle density.}. In spite of the
differences in the colliding systems, results for various centralities in
U+U collisions almost trace the ones in Au+Au collisions, which would
suggest existence of scaling behavior in
$v_2/\varepsilon_{\mathrm{part}}$ versus $(1/S)dN_{\mathrm{ch}}/d\eta$.

However, the behavior of $v_2/\varepsilon_{\mathrm{part}}$ in
  Pb+Pb collisions at $\sqrt{s_{NN}} =$ 2.76 TeV is very different.
  In these collisions the system is
  much denser than in the collisions at RHIC energies. The maximum
  temperature
  at the initial time $\tau_{0}$ = 0.6  fm/$c$ is $T_{0}=474$, 456, 436 MeV
  ($e_{0}=96.2$, 81.7, 67.8 GeV/fm$^{3}$)
  for $dN_{\mathrm{ch}}/d\eta \sim 1600$, 1400, 1200
  in 5\% most central collisions, respectively,
which corresponds to roughly 2--2.5 times larger transverse
  density than in most central Au+Au collisions at $\sqrt{s_{NN}} =
  200$ GeV.
As can be seen, $v_2/\varepsilon_{\mathrm{part}}$ no longer
  follows the scaling curve seen at the top RHIC energy, but it
  reaches $\sim 0.26$--$0.3$ in central collisions. This value is $\sim
  20$--$35\%$ larger than the value at RHIC, which is often considered a
  hydrodynamical upper limit in the literature~\cite{Voloshin:2008dg}.

 \begin{figure}[htb]
\includegraphics[width=3.4in]{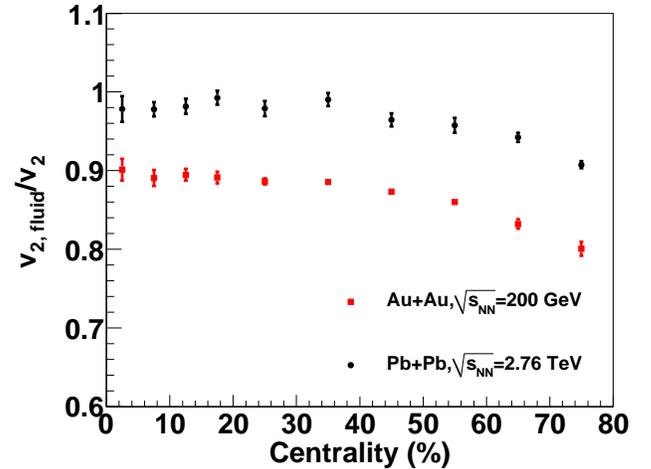}
\caption{The ratio of $v_2$ generated during the
  hydrodynamical evolution to the final \vtwo,
  $v_{2,\mathrm{fluid}}/v_2$, in Au+Au and Pb+Pb collisions at
  $\sqrt{s_{NN}} = 200$ GeV and 2.76 TeV, respectively.}
 \label{fig:v2fluid}
 \end{figure}

  The reason for this breaking of the scaling is not in the cascade
  treatment of the hadronic phase. If anything, dissipation should
  reduce $v_2$, so hadron cascade cannot be responsible for the large
  value of $v_2/\varepsilon_{\mathrm{part}}$ seen here. We have also
  checked that at LHC, the major part of $v_2$ is generated during the
  hydrodynamical stage, and the effects of hadronic cascade are less
  important than at RHIC. The ratio of \vtwo generated during the
  hydrodynamical evolution to the final $v_2$,
  $v_{2,\mathrm{fluid}}/v_2$, in collisions at RHIC and LHC is shown
  in Fig.~\ref{fig:v2fluid}. As can be seen, the contribution of
  hadronic cascade to the total \vtwo with default setting at LHC is less than 5\% at most
  centralities. On the other hand, the contribution reaches 10-20\% of
  the total $v_2$ in Au+Au collisions at the top RHIC energy.

  To further study the collision energy dependence of
  $v_2/\varepsilon_{\mathrm{part}}$, we do the calculation using the
  lower RHIC energy $\sqrt{s_{NN}} =$ 62.4 GeV. As seen in
  Fig.~\ref{fig:v2eccLHC}, the ratio in central collisions (0-30\%)
  deviates from the
scaling curve seen at $\sqrt{s_{NN}} =$ 200 GeV,
but the amount of the deviation might be too small
to be experimentally observable.
The collision energy independence of 
$v_2/\varepsilon_{\mathrm{part}}$
is seen in $(1/S)dN_{\mathrm{ch}}/d\eta \lsim 10$ (fm$^{-2}$)
which corresponds to $dN_{\mathrm{ch}}/d\eta \lsim 50$
where hadronic cascading plays a major role
in the whole evolution.
Note that the collision energy dependence of 
$v_2/\varepsilon_{\mathrm{part}}$
is consistent with the early calculations
where $v_2$ continuously increases
with total pion multiplicity $dN_{\pi}/dy$ at midrapidity
up to 3000 at a fixed impact parameter ($b=7$ fm) \cite{Kolb:2000sd}.
See also Ref.~\cite{Hirano:2007xd} for a previous calculation
in which the bag model equation of state
and a higher switching temperature $T_{\mathrm{sw}}=169$ MeV
were used.


Recently, a prediction of elliptic flow 
as a function of transverse charged particle density
up to the LHC energies was made 
using viscous hydrodynamics
in Ref.~\cite{Luzum:2009sb}.
To avoid the uncertainties associated with the freeze-out process,
\vtwo was evaluated in that paper by calculating
the momentum anisotropy 
\begin{eqnarray}
e_{p} & = & \frac{\int dx dy(T^{xx}-T^{yy})}{\int dx dy (T^{xx}+T^{yy})}
\end{eqnarray}
and relying on an empirical formula
$v_{2} \approx e_{p}/2$ \cite{Kolb:1999it}.
We have checked the validity of this formula in our calculations,
  and found that in the collisions at the LHC energy, the ratio is
  rather $v_2/e_p \approx 2/3$, not 1/2. This discrepancy is not
  surprising. First, it is known that the ratio strongly depends on
  the freeze-out temperature~\cite{Huovinen:2003fa}. The momentum
  anisotropy depicts the anisotropy of the collective motion, whereas
  $v_2$ reflects the anisotropy of the momenta of individual
  particles, which includes thermal motion, the effects due to
  resonance decays \cite{Hirano:2000eu} and to the shape of the
  source~\cite{Huovinen:2001wn}. Second, the formula was found to hold
  in ideal fluid calculations. How dissipation affects it cannot be
  calculated, but has to be observed on case by case basis.

To summarize, we predicted elliptic flow parameter $v_{2}$
in U+U collisions at $\sqrt{s_{NN}} = 200$ GeV and
in Pb+Pb collisions at $\sqrt{s_{NN}} = 2.76$ TeV
using a hybrid approach
which combines ideal hydrodynamic
description of the QGP fluid
and kinetic description of the hadronic gas.
Due to deformation of uranium,
eccentricity is larger in U+U collisions
than in Au+Au collisions.
We found the maximum transverse particle density
is $\sim$ 6\% larger 
in 0-5\% central U+U collisions.
$v_2/\varepsilon_{\mathrm{part}}$
in U+U collisions
follows the results in Au+Au collisions,
which suggests a scaling behavior
between $v_2/\varepsilon_{\mathrm{part}}$
and $(1/S)dN_{\mathrm{ch}}/d\eta$.
However, at the LHC energy, $v_{2}/\varepsilon_{\mathrm{part}}$
does \emph{not} follow the same scaling curve
 and reaches the maximum value of
$\sim 0.26$--$0.30$ depending on the final particle multiplicity.
This is clearly larger than the corresponding maximum value at
the top RHIC energy, $v_{2}/\varepsilon_{\mathrm{part}} \sim$ 0.22,
and the so-called hydrodynamic limit for $v_2/\varepsilon$ is not the
same at RHIC and LHC energies.

\acknowledgments
The work of T.H. (Y.N.) was partly supported by
Grant-in-Aid for Scientific Research
No.~22740151 (No.~20540276).
T.H. is also supported under
Excellent Young Researchers Oversea Visit Program
(No.~21-3383)
by Japan Society for the Promotion of Science.
P.H.'s work is supported by the ExtreMe Matter Institute (EMMI).
We acknowledge fruitful discussion with A.~Dumitru.
T.H. thanks members in the nuclear theory group
at Lawrence Berkeley National Laboratory
for a kind hospitality during his sabbatical stay
and M.~Gyulassy for his suggestion to calculate
$v_2$ at the LHC energies.


\end{document}